\documentclass[conference,letterpaper]{IEEEtran}
\IEEEoverridecommandlockouts
\usepackage{cite}
\usepackage{amsmath,amssymb,amsfonts,bm}
\usepackage{algpseudocode}
\usepackage{algorithm,float}
\usepackage{graphicx,subfigure}
\usepackage{textcomp}
\usepackage{xcolor}
\usepackage{multirow}
\usepackage{balance} 
\usepackage{colortbl}
\usepackage{url}
\usepackage{caption}
\usepackage{graphicx}
\usepackage{grffile}
\def\BibTeX{{\rm B\kern-.05em{\sc i\kern-.025em b}\kern-.08em
T\kern-.1667em\lower.7ex\hbox{E}\kern-.125emX}}

\definecolor{LightCyan}{rgb}{0.88,1,1}
\definecolor{LightGray}{rgb}{0.88,0.88,0.88}

\newtheorem{rem}{Remark}
\definecolor{org}{rgb}{0.9,0.4,0.1}

\begin{document}
\title{Interference-Aware Constellation Design for Z-Interference Channels with Imperfect CSI}

\author{Xinliang Zhang$^\dagger$, Mojtaba Vaezi$^\dagger$, and Lizhong Zheng$^\ddagger$ 
}

\author{ 
	\IEEEauthorblockN{Xinliang Zhang$^\dagger$, Mojtaba Vaezi$^\dagger$, and Lizhong Zheng$^\ddagger$}
	\IEEEauthorblockA{
		$^\dagger$Department of Electrical and Computer Engineering, 
		Villanova University, Villanova, PA 19085, USA\\
		$^\ddagger$EECS Department, Massachusetts Institute of Technology, Cambridge, MA 02139, USA\\
		Emails:  \{xzhang4, mvaezi\}@villanova.edu$^\dagger$,
	 lizhong@mit.edu$^\ddagger$}
}

\maketitle

\begin{abstract}
A deep autoencoder (DAE)-based end-to-end communication over the two-user 
Z-interference channel (ZIC) with finite-alphabet inputs is designed in this paper. The design is for
imperfect channel state information (CSI) where both estimation and quantization 
errors exist.
The proposed structure jointly optimizes the 
encoders and decoders to  
generate interference-aware constellations that adapt their shape to the interference intensity 
in order to minimize the bit error rate. A normalization layer is designed to guarantee 
an average power constraint in the DAE while allowing the architecture to generate constellations with 
nonuniform shapes. This brings further shaping gain compared to standard uniform constellations such 
as quadrature amplitude modulation.     
The performance of the  DAE-ZIC is compared with two conventional methods, i.e., standard and 
rotated constellations. The proposed structure significantly enhances the performance of the 
ZIC. Simulation results confirm bit error rate reduction in all interference regimes (weak, 
moderate, and strong). At a signal-to-noise ratio of $\rm 20 dB$, the improvements reach about two
orders of magnitude when only quantization error exists, indicating
that the DAE-ZIC is highly robust to the interference compared
to the conventional methods.

\end{abstract}


%
\section{Introduction}
Interference is a central issue in today's \textit{multi-cell} networks.
The information-theoretic model for a multi-cell network is the \textit{interference channel} (IC). 
There have been many efforts to find the capacity of the IC either 
with the same generality and accuracy used by Shannon for point-to-point systems 
\cite{carleial1975case,han1981new} or 
by seeking approximate solutions with a guaranteed gap to optimality at any signal-to-noise
ratio (SNR) \cite{etkin2008gaussian}.
However, the capacity region of the two-user IC is only  known for strong  
interference   where decoding and canceling the 
interference is optimal \cite{carleial1975case}. 
Also, at very 
weak interference,  sum-capacity is achievable by treating interference as noise 
\cite{motahari2009capacity}.
In general,  Han-Kobayashi encoding is the best achievable scheme \cite{han1981new}, which 
decodes part of the interference  and treats the 
remaining as noise.

The aforementioned Shannon-theoretic works are based on  Gaussian inputs.  Despite being  
theoretically optimal, Gaussian alphabets are continuous and unbounded, and thus, are  rarely 
applied in real-world communication. 
In practice, signals are  generated using finite alphabet sets, 
such as phase-shift keying (PSK) 
and  quadrature amplitude modulations (QAM). 
The performance gap between the finite alphabet input  and the Gaussian input design is 
non-negligible \cite{wu2013linear}. However, conventional finite-alphabet approaches are 
based 
on predefined uniform constellations like QAM. 
These constellations are defined for point-to-point systems 
\cite{foschini1974optimization,goldsmith1997variable,barsoum2007constellation} 
 and their constellation shaping is oblivious to interference. Such an inability to respond to 
interference is an obstacle to improving the bit-error rate and spectral efficiency of today’s 
interference-limited communication systems.



In this paper, we consider the two-user  one-sided IC,   also known as the Z-interference channel (ZIC) \cite{vaezi2016simplified},
with imperfect CSI. 
Previous works have examined the ZIC with finite alphabet inputs and uniform constellations in certain regimes.  In 
\cite{knabe2010achievable}, it is shown that rotating one input constellation (alphabet) can 
improve the sum-rate of the two-user IC in strong/very strong interference 
regimes. Later, an exhaustive search for finding the optimal rotation of the signal 
constellation was presented in \cite{ganesan2012two}.   
The focus of the above papers is to maximize the achievable rates, and they do not study 
bit-error rate (BER) performance. BER is a critical metric, and interference can severely 
increase the BER by distorting the received constellation when uniform constellations like QAM 
are employed.

Deep autoencoder (DAE)-based end-to-end  communication is an emerging approach to  
finite-alphabet communication in which BER is the main performance measure and constellation 
design is inherent to it.  Various groups have proposed DAE-based communication both for single- and multi-user systems
\cite{o2017introduction,song2020benchmarking,zhang2021svd}. 
Particularly,  \cite{o2017introduction, erpek2018learning, wu2020deep} have studied 
communication over the IC. These works, however, are only for the symmetric interference case 
and compare their results with simple baselines (e.g., quadrature phase shift keying (QPSK)), but 
we know QPSK performs much poorer than a rotated
QPSK \cite{knabe2010achievable,ganesan2012two}.
In addition, those structures assume perfect knowledge of channel state information (CSI), but 
the extension from perfect CSI to imperfect CSI  is not straightforward and has not been 
explored.

This paper sheds light on DAE-based communication over asymmetric interference with both 
perfect and imperfect CSI.  Specifically, we design and train novel DAE-based architectures  for 
the ZIC with finite-alphabet inputs.  
In this architecture, we have two transmitter-receiver DAE pairs that work together to mitigate interference and adapt the constellation to the interference intensity, leading to improved BER. Key 
contributions of the paper are as follows: 
\begin{itemize}
\item We design a DAE-based transmission structure for the ZIC which works both for imperfect and  
perfect CSI at different interference regimes, including weak, moderate, and strong 
interference. 
In the proposed architecture, we have designed an average power constraint normalization layer to
allow generating \textit{nonuniform constellations} to use the in-phase and 
quadrature-phase (I/Q) plane efficiently.  
The resulting constellations are adaptive to 
the interference intensity and morph in a way that the receivers see distinguishable symbols. 
\item This is the first study examining finite-alphabet ZIC
under imperfect CSI. We consider both estimation and quantization errors. 
Particularly, the CSI
estimation error confuses the DAE and brings difficulty in training and testing 
performance. The quantization error, coming from the limited feedback channel capacity,  is an 
additional error that brings unwanted rotations to the constellations. 
To overcome these challenges, we first simplify the CSI parameters by designing an equivalent system model and then we build the DAE to reduce the BER. 
\end{itemize}

 For benchmarking, we use \textit{rotated} uniform constellations  which are more 
competitive than unrotated constellations.  The  proposed DAE-ZIC shows significantly better 
BER performance for all interference regimes (weak, 
moderate, and strong interference). The overall BER reduction is about 40\%, and the gap between 
the DAE-ZIC and the best conventional method is even larger when quantization error exists.   

We organize the remainder of this paper as follows. We first elaborate on the 
ZIC system model and its simplification in Section~\ref{sec_impCH}. 
We next introduce the DAE design and the training approach in Section~\ref{sec_dae}. 
We present the simulation results and conclusions in  Sections~\ref{sec_simulation} and \ref{sec_con}, 
respectively.

\section{System Model with Imperfect CSI}\label{sec_impCH}
\subsection{Channel Model of the ZIC}
\begin{figure}[tb] 
	\centering
	\includegraphics[scale=.25, trim=0 0 0 0 , 
		clip]{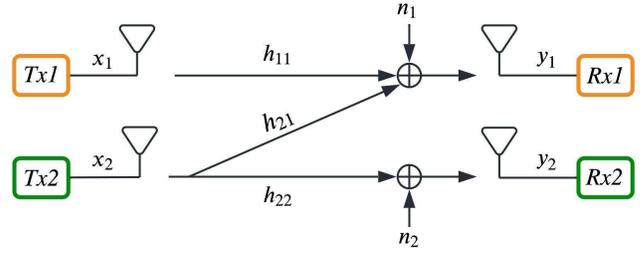}
	\caption{Original system model of the ZIC where 	$h_{ij}=r_{ij}e^{j\theta_{ij}}$.}
	\label{fig_sys_impCH}
\end{figure}

The system model of two-user single-input 
single-output ZIC is shown in Fig.~\ref{fig_sys_impCH}. The two 
transmitter-receiver pairs  wish to reliably transmit their messages while the transmission of the first pair is interfered with by the second one. The four nodes are named \textit{Tx1}, \textit{Tx2}, \textit{Rx1}, and \textit{Rx2}.  The 
received signals at \textit{Rx1}, and \textit{Rx2} can be written as
\begin{subequations}\label{eq_ZIC_sys}
	\begin{align}
		&y_1 = h_{11}x_1 + h_{21}x_2 + n_1,\\
		&y_2 = h_{22}x_2  + n_2,
	\end{align}
\end{subequations}
in which $x_i$ and $y_i$, $i\in\{1,2\}$, denote the transmitted and received symbols,  $n_{i}$ is 
white Gaussian noise with mean zero and variance $\sigma_i^2$, and actual channel coefficients are given by
\begin{align}
	h_{ij}\sim\mathcal{CN}(\mu_H, \sigma_H^2), \;i,j\in\{1,2\},\label{eq_perfectCH}
\end{align}
where $\mu_H$ and $\sigma_H^2$ are the mean and variance of the channel. 
$h_{12}=0$ by the definition of the ZIC. The interference intensity is defined as 
\begin{align}
\alpha\triangleq\left|{h_{21}}{h_{11}^{-1}}\right|^2.
\end{align}
In practice, perfect channel gains are not available. 
{${h}_{11}$ and ${h}_{21}$ are estimated by \textit{Rx1} whereas ${h}_{22}$ is estimated by \textit{Rx2}.} The
CSI imperfectness comes from two sources: {1) the error in the CSI estimation at the receivers' side and 2) the quantization error when feeding CSI back to the transmitters.} 
We give the details of two types of errors as follows. 

\subsubsection{Estimation Errors}
The estimated channel coefficients are determined by the actual channel and 
estimation error, which is modeled as \cite{chen2005performance} 
\begin{align}\label{eq_impCH_org} 
	\hat{h}_{ij}=\hat{r}_{ij}e^{j\hat\theta_{ij}}\triangleq{h}_{ij}-\varepsilon_{ij},\; i,j\in\{1,2\},
\end{align}
where $\hat{r}_{ij}$ and $\hat\theta_{ij}$ are amplitude and phase of $\hat{h}_{ij}$,   $h_{ij}$  is the actual channel, and 
\begin{align}
	\varepsilon_{ij}\sim\mathcal{CN}(0,\sigma_E^2)\label{eq_estErrVar}
\end{align}
is the estimation error with the variance $\sigma_E^2$. 
Hence, we have $\hat{h}_{ij}\sim\mathcal{CN}(\mu_H,\sigma_H^2+\sigma_E^2)$. 
The imperfectness of $h_{ij}$ affects the decoding process.

\subsubsection{Quantization Errors}
The transmitters require the knowledge of CSI in closed-loop systems. 
However, due to the limited feedback resources, the feedback information is quantized with reduced accuracy.
Thus, quantization brings in another imperfectness. {For example, \textit{Rx1} estimates ${h}_{11}$ and ${h}_{21}$ and gets the estimated interference intensity, 
 $\hat{\alpha}=|\hat{h}_{21}\hat{h}_{11}^{-1}|^2$. 
To let all four nodes access $\hat{\alpha}$, a quantized value of that 
\begin{align}
	&\alpha_q=Q(\hat\alpha),\label{eq_fb_alp}
\end{align} 
 is fed back to \textit{Tx1}, \textit{Tx2}, and \textit{Rx2},  where  $Q(\cdot)$ is a  quantizer of its input variable.  $Q(\cdot)$ uniformly divides the considered  range of $\hat{\alpha}$, which is $[0,3]$, into $2^{N_q}$ segments. The middle value of the segment is the quantization result}.

\subsection{Simplified Model of the ZIC}
\begin{figure}[tb] 
	\centering
	\includegraphics[scale=.24, trim=0 0 0 0 , 
	clip]{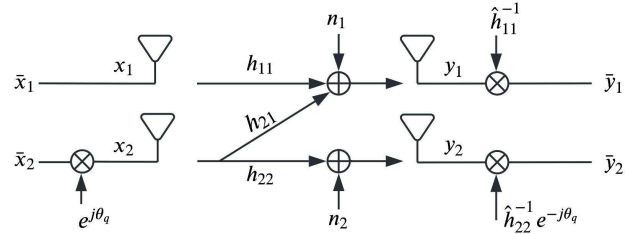}
	\caption{Pre- and post-processed model of the ZIC to simplify it (see Remark~\ref{rem:ZIC}) for a DAE-based implementation. }
	\label{fig_sys_eq}
\end{figure}
\begin{figure*}[tbp] 
	\centering
	\includegraphics[scale=.37, trim=0 5.6 0 8.1, 
	clip]{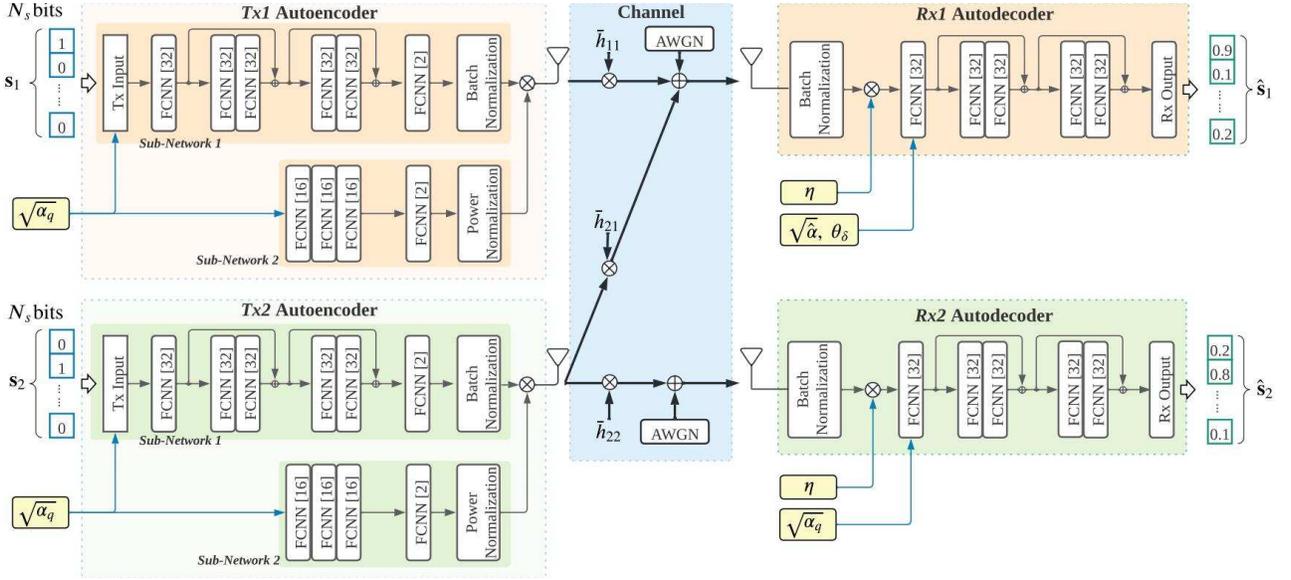}
	\caption{The architecture of the two-user  DAE-ZIC implemented by two pairs of 
		deep autoencoders. Each transmitter of the ZIC contains two sub-networks. 
		\textit{Sub-network 1} mainly generates the constellation and \textit{sub-network 2} is 
		used to implement the average power constraint. The receivers decode their bits from the 
		received signal.  $\eta$ is 	a power control parameter in
		\eqref{eq_eta}.}
	\label{fig_DAE_net} 
\end{figure*}
In this subsection, we simplify the above model for DAE. 
Such a model shown in Fig.~\ref{fig_sys_eq} is achieved by pre- and post-processing  in \textit{Tx2}, \textit{Rx1}, and \textit{Rx2}.
Specifically, \textit{Tx2} applies pre-processing by multiplying $e^{j\theta_q}$ to the source signal defined as $\bar{x}_2$, i.e., $x_2=\bar{x}_2 e^{j\theta_q}$. $\theta_q$ is estimated by \textit{Rx1} and fed back to other nodes designed as  
\begin{align}
	&\theta_q\triangleq Q(\hat\theta_{11}-\hat\theta_{21})=(\hat\theta_{11}
	-\hat\theta_{21})+\theta_\delta,\label{eq_fb_alg}
 \end{align}
where $\hat\theta_{11}$ and $\hat\theta_{21}$ are defined in   \eqref{eq_impCH_org}, and   $\theta_\delta$ is the quantization error.
Correspondingly, ${x}_1=\bar{x}_1$ since $h_{12}=0$.
The received signals after  post-processing are
\begin{subequations}
\begin{align}
	&\bar{y}_1
	=\bar{h}_{11}\bar{x}_1+\bar{h}_{21}\bar{x}_2+\bar{n}_1, \label{eq_y_prime_1}\\
	&\bar{y}_2=
	\bar{h}_{22}\bar{x}_2+\bar{n}_2,\label{eq_y_prime_2}
\end{align}
\end{subequations}
where  the equivalent channels are given by
\begin{subequations}
	\begin{align}
		&\bar{h}_{ii} \triangleq {h}_{ii}\hat{h}_{ii}^{-1} = 1+\varepsilon_{ii} \hat{h}_{ii}^{-1},\;\label{eq_impCh_dir} \\
		&\bar{h}_{21} \triangleq{h}_{21}\hat{h}_{11}^{-1}e^{j\theta_q} =   \sqrt{\hat{\alpha}}e^{j\theta_\delta}+\varepsilon_{21} 
		\hat{h}_{11}^{-1}e^{j\theta_q},\label{eq_impCh_cross}
	\end{align}
\end{subequations}
and  $\bar{n}_i$ is  the equivalent noise which is  given by
\begin{align}
		& \bar{n}_i\triangleq n_i\hat{h}_{ii}^{-1}\sim\mathcal{CN}(0,\sigma_N^2 
\hat{r}_{ii}^{-2}). \label{eq_noisePw}
\end{align}

\addtolength{\topmargin}{0.04in}

\begin{rem} \label{rem:ZIC}
	If the estimation error and the quantization error are absent, i.e., $\varepsilon_{ij}=0$ and $\theta_\delta =0$, the system in \eqref{eq_y_prime_1}-\eqref{eq_y_prime_2} reduces to the one with perfect CSI case  \cite{vaezi2016simplified, kramer2004review,zhang2023wcncIC}, in which the channel gains of the direct and interference links become  $\bar{h}_{ii}=1$ and $\bar{h}_{21}=\sqrt{\alpha}$. These are all real-valued.
	\end{rem}


\section{Deep Autoencoder for ZIC for Imperfect CSI}\label{sec_dae}
Existing studies \cite{knabe2010achievable, 	ganesan2012two} use standard QAM constellations for interference channels. Such constellations have fixed symbols and are not adjustable according to the  
interference intensity. To improve the transmission performance, we propose a 
DAE-based transmission for 
the two-user ZIC, named DAE-ZIC. We show the architecture in Fig.~\ref{fig_DAE_net}.

\subsection{The Architecture of DAE-ZIC}

\subsubsection{Network Input}
Each transmitter sends $N_s$ bits to the corresponding receiver.  The feedback of the 
interference intensity $\sqrt{\alpha_q}$  is  appended to the input bit vector.  The two 
transmitters are expected to jointly design their constellations and the 
receivers will decode correspondingly.

\subsubsection{Transmitter DAE}
As shown in Fig.~\ref{fig_DAE_net}, the DAE in each transmitter contains two 
sub-networks. 
\textit{Sub-network~1} converts the input bit-vector to symbols that take the value of 
$h_{21}$ into consideration. \textit{Sub-network~2} performs power allocation, which controls 
the power of the I/Q components. 

The main components of \textit{sub-network 1} are are fully connected neural networks
(FCNN), residual connections, and  the output batch 
normalization (BN) layer.\footnote{The FCNN and residual connections  inherit the design of 
the  
point-to-point MIMO transmission in \cite{zhang2021svd}.}  The activation function of the FCNN 
layers is \textit{tanh} except for the last 
layer, which has two hidden nodes and no activation function.  Assume the batch size is $N_B$, 
and 
the output of the last FCNN is 
	$\mathbf{X}_{\textmd{fcnn}}\triangleq [\mathbf{x}_{\textmd{fcnn}}^{\textmd{I}}, \;
	\mathbf{x}_{\textmd{fcnn}}^{\textmd{Q}}]$,
where $\mathbf{x}_{\textmd{fcnn}}^{\textmd{I}}$ and  
$\mathbf{x}_{\textmd{fcnn}}^{\textmd{Q}}\in\mathbb{R}^{N_B\times 1}$ are the outputs of the 
two hidden nodes and  represent I/Q of the complex-valued signal.

Since the FCNN has unbounded outputs, it cannot guarantee a power constraint at the 
transmitter. We propose a transmitter architecture as shown in Fig.~\ref{fig_DAE_net} to achieve an 
average power constraint at each 
antenna. First, we  use BN in \textit{sub-network 1}  to unify  the 
average power of I/Q independently.  
The BN layer linearly normalizes  
${\mathbf{x}}_{\textmd{fcnn}}^{\textmd{I}}$ and ${\mathbf{x}}_{\textmd{fcnn}}^{\textmd{Q}}$, in 
which the
normalized vectors  
$\mathbf{x}_B^{\textmd{I}}$ and $\mathbf{x}_B^{\textmd{Q}}$ are
\begin{align}
\mathbf{x}_B^{\textmd{I}}\triangleq\beta^{\textmd{I}} {\mathbf{x}}_{\textmd{fcnn}}^{\textmd{I}},
\; \text{and} \quad 
\mathbf{x}_B^{\textmd{Q}}\triangleq\beta^{\textmd{Q}} {\mathbf{x}}_{\textmd{fcnn}}^{\textmd{Q}},
\end{align}
where $\bm{\beta}\triangleq[\beta^{\textmd{I}},\;\beta^{\textmd{Q}}]^T$ contains two factors 
for 
normalization. 
Then, the powers of $\mathbf{x}_B^{\textmd{I}}$ and $\mathbf{x}_B^{\textmd{Q}}$ 
are modified by \textit{sub-network 2}.
\textit{Sub-network 2} has two output values:  
$\gamma^{\textmd{I}}$ and  $\gamma^{\textmd{Q}}$.  {The FCNN layers in 
\textit{sub-network 2} determine the power allocated to the I/Q components  based on the 
input 
value 
$\sqrt\alpha$. The power normalization (PN) block in \textit{sub-network 2} limits the total 
power 
 to $P_t$.} Defining 
${\bm\gamma}\triangleq[\gamma^{\textmd{I}}, 
\;\gamma^{\textmd{Q}}]^T\in\mathbb{R}^{2\times1}$, we should have
${\bm\gamma}^T{\bm\gamma}=P_t$. 
Finally, the outputs of the BN and PN are multiplied together, 
i.e.,
\begin{align}
\mathbf{x}_\textmd{out}^{\textmd{I}} \triangleq 
\gamma^{\textmd{I}} \mathbf{x}_B^{\textmd{I}},\; \text{and} \quad 
\mathbf{x}_\textmd{out}^{\textmd{Q}} \triangleq 
\gamma^{\textmd{Q}} \mathbf{x}_B^{\textmd{Q}}.
\end{align}
The powers of $\mathbf{x}_\textmd{out}^{\textmd{I}}$ and 
$\mathbf{x}_\textmd{out}^{\textmd{Q}}$  are 
$\gamma^{\textmd{I}}$ and  $\gamma^{\textmd{Q}}$, respectively. 
To summarize, BN is applied to the I/Q components  along the time, while PN normalizes the I/Q components at each time. Hence, the 
two normalization operations are implemented in different dimensions. In this way, the 
average power constraint is reached.


\subsubsection{Channel Implementation}
The channels are implemented by FCNN layers independently. 
The weight,  $\bar{\mathbf{H}}_{ij}$, is  the real form of the  channel 
$\bar{h}_{ij}=\bar{h}_{ij}^{\textmd{I}}+j\bar{h}_{ij}^{\textmd{Q}}$ in 
\eqref{eq_impCh_dir}-\eqref{eq_impCh_cross}, 
\begin{align}\label{eq_net_impCH}
	\bar{\mathbf{H}}_{ij}=\left[
	\begin{matrix}
		\bar{h}_{ij}^{\textmd{I}}& -\bar{h}_{ij}^{\textmd{Q}}\\
		\bar{h}_{ij}^{\textmd{Q}}& \quad\bar{h}_{ij}^{\textmd{I}}
	\end{matrix}
	\right].
\end{align}
These channel layers have zero-bias, no activation function, and are non-trainable.

\subsubsection{Receiver DAE}
The received signals are $\bar{y}_1$ and $\bar{y}_2$.  To ensure the receiver 
networks have a finite input range, we use BN layers to 
unify the power of the received signals, i.e.,
\begin{align}\label{eq_rx_BN}
{y}_{\textmd{B},i} = \xi\bar{y}_i,\;\; \mathbb{E}\{|{y}_{\textmd{B},i}|^2\} = 1,\;\; \forall i\in\{1,2\},
\end{align}
where $\xi$ is a coefficient to reach the  unit power. The process details and settings are 
the same as the ones in the transmitter.

We further define the \textit{desired signal} for \textit{Rx1} as
$x_{\textmd{D},1} \triangleq \bar{x}_1+\sqrt{\alpha}\bar{x}_2$
which contains the true desired signal $\bar{x}_1$ and the interference 
$\sqrt{\alpha}\bar{x}_2$. The goal of the receiver is to decode $x_1$ for an arbitrary  $\bar{x}_2$ in its 
constellation. The \textit{desired signal} of \textit{Rx2} is  
$x_{\textmd{D},2} \triangleq \bar{x}_2$.
However, the normalization of the received signal \eqref{eq_rx_BN} makes the power of the 
\textit{desired signal} vary with the SNR. Hence, the autoencoder should adjust the decoding 
boundary according to the SNR, which is an extra burden. 
So, we turn to normalize the desired signal using a linear factor, $\eta$,  multiplied 
by the batch normalization output, i.e.,
\begin{align}\label{eq_eta}
y_{\textmd{D},i} = \eta \cdot {y}_{\textmd{B},i},\;
\eta\triangleq\sqrt{1+{P_{\textmd{D,i}}}{\sigma_N^{-2}}},\quad \forall i\in\{1,2\},
\end{align}
where $P_{\textmd{D},i}$ is the power of the \textit{desired  signal} $x_{\textmd{D},i}$ 
and $\sigma_N^2$ is the noise power. In short, the BN normalizes the 
\textit{desired  signals} using pre-processing $\eta$.

Besides, the receivers append the available parameters as additional input to the FCNN layer.
Since the estimated interference intensity $\sqrt{\hat{\alpha}}$ and the quantization error of 
feedback angle $\theta_\delta$ are known at \textit{Rx1}, we input these parameters to the 
autoencoder of \textit{Rx1}. For \textit{Rx2}, the feedback of the interference intensity 
$\sqrt{\alpha_q}$  is  the additional input.
The final output of the DAE is an estimation of the transmitted bit-vectors, $\hat{\mathbf{s}}_1$ 
and $\hat{\mathbf{s}}_2$, as shown in Fig.~\ref{fig_DAE_net}. The output layer uses soft-max. 
More specifically, the activation function is sigmoid. 
\begin{algorithm}[tbp]
	\caption{Training Procedure  for the DAE-ZIC}\label{alg_Train}
	\begin{algorithmic}[1]
		\State Inputs: $N_s$, $\alpha_{\min}$,  $\alpha_{\max}$, $\mu_H$, $\sigma_H^2$, 
		$\sigma_E^2$, $T$, and $N_q$.
		\State Set $P_t=1$W, SNR$=10$dB, $N_\alpha=30,000$, $E_p=10$, $N_B=10^4$, and  
		$l_r=10^{-2}$ which will  drop to $d_rl_r=0.95l_r$ after 
		every $N_d=200$ trained channels.
		\State Initialize the DAE-ZIC network.
		\For {index $i_\alpha$ from $1$ to $N_\alpha$}
		\State Uniformly and randomly select one $\alpha\in[\alpha_{\min}, \alpha_{\max}]$.
		\While{1}
		\State Randomly generate $h_{11}$ and $h_{22}$ using \eqref{eq_perfectCH}.
		\State Randomly generate $\varepsilon_{11}$, 
		$\varepsilon_{22}$, and $\varepsilon_{21}$ using \eqref{eq_estErrVar}.
		\State \textbf{If} \eqref{eq_impCH_thres} satisfied, \textbf{then} Break.
		\EndWhile
		\State Uniformly generate  $\Delta\theta_q$ in $\frac{1}{2^{N_q}}[-\pi, \pi]$.
		\State Update  $\alpha_q$, $\theta_\delta$, and $\theta_q$ using \eqref{eq_fb_alp} and 			\eqref{eq_fb_alg}. 
		\State Normalize the channels using \eqref{eq_impCh_dir} and \eqref{eq_impCh_cross}.
		\For {index $i_e$ from $1$ to $E_p$}
		\State {Randomly  generate $N_B$ bit vectors.}
		\State Update the weights of the DAE-ZIC  using Adam. 
		\EndFor
		\State Set learning rate $l_r=d_r l_r$ if $i_\alpha/N_d$ is an integer.
		\EndFor
	\end{algorithmic}
\end{algorithm} 
\subsubsection{Loss Function}
In our DAE-ZIC, each receiver has its own estimation of the transmitted bits. Then, the  overall 
loss 
function of the DAE-ZIC is 
$\mathcal{L}=\mathcal{L}_1+\mathcal{L}_2$,
where $\mathcal{L}_1$ and $\mathcal{L}_2$ are the losses  at \textit{Rx1} and \textit{Rx2}. 
In this 
paper, we use binary cross-entropy as the loss function, i.e.,
\begin{align}
\mathcal{L}_i=\frac{1}{N_B}\sum_{n=1}^{N_B}    \mathbf{s}_{i,n}^T\log\hat{\mathbf{s}}_{i,n} +
(1-\mathbf{s}_{i,n})^T\log(1-\hat{\mathbf{s}}_{i,n}),
\end{align}
where $i\in\{1,2\}$ distinguishes  the users,  $N_B$ is the batch size, $\mathbf{s}_{i,n}$ is the 
$n$th  
input bit-vector in the batch, and $\hat{\mathbf{s}}_{i,n}$ is corresponding the output. The 
loss function treats each element of the DAE output as a zero/one classification task.  
Cross-entropy is used  to evaluate each classification task. Finally, the loss is 
the summation of the loss of $N_s$ tasks, where $N_s$ is the number of bits in the transmission.
In the training process, the back propagation algorithm passes $\mathcal{L}_1$ to \textit{Rx1} 
and this will further go to \textit{Tx1} and \textit{Tx2}. The  $\mathcal{L}_2$  affects the 
\textit{Rx2} and \textit{Tx2}.

\subsection{Training Procedure of the DAE-ZIC}\label{sec_train}
We use separate instances of DAEs for different ranges
of the interference gain $\alpha$. This is because training one network over all values of 
$\alpha$ is not easy.   
In each training, we select the $N_s$  and the desired range for 
$\alpha\in[\alpha_{\min}, \alpha_{\max}]$. We train the DAE repeatedly using random values of 
$\alpha$ in this 
interval. For each $\alpha$, the DAE is trained through epochs $E_p$,  mini-batch size $N_B$, 
and a constant  learning rate $l_r$. After training the DAE for $N_d$ different values of $\alpha$, 
the 
learning rate is reduced to $d_r l_r$.
The detailed training procedure is summarized in Algorithm~\ref{alg_Train}. 
Also, we choose the best DAEs out of five individually trained networks with the same hyper 
parameters. The best is defined by the average loss on ten randomly generated values of $\alpha$ in 
$[\alpha_{\min}, \alpha_{\max}]$. To avoid the numerical problem, in both training and testing, we only use channels 
satisfying
\begin{align}\label{eq_impCH_thres}
	\max(\left|{\varepsilon_{11}}{\hat{h}_{11}^{-1}}\right|,  
	\left|{\varepsilon_{22}}{\hat{h}_{22}^{-1}}\right|,  
	\left|{\varepsilon_{21}}{\hat{h}_{11}^{-1}}\right|) <  T,
\end{align}
where $T$ is a threshold. $T$ is set as one in this paper so that the estimation errors are not dominant in $\bar{h}_{ii}$ in \eqref{eq_impCh_dir}.


{
\section{Performance Analysis}\label{sec_simulation}
The performance  is evaluated and compared for the three methods listed below. To be fair to the 
users, we the use maximum (worst) BER of the two as the measurement. 
\begin{itemize}
	\item \textit{DAE-ZIC}: The proposed method which designs nonuniform constellations based on the interference intensity.
	\item \textit{Baseline-1}: The transmitters directly use standard QAM. 
	\item \textit{Baseline-2}: \textit{Tx1} uses standard QAM, while \textit{Tx2} rotates 
	the standard QAM symbols based on the interference intensity \cite{ganesan2012two}.
\end{itemize}


The implementation of DAEs are performed in TensorFlow
and the baselines are performed in MATLAB.

\subsection{CSI with Estimation Error}
The received constellations at \textit{Rx1} generated by the baselines and the 
proposed DAE-ZIC are shown in Fig.~\ref{fig_const}. {In this simulation,} we set $N_s=2$, i.e.,
each user has $2^{N_s}=4$ information symbols. The estimated ${\hat\alpha}$ and 
the actual equivalent channels are given in Table~\ref{tab_ch}.  The transmit power is unity, 
and the SNR is 10dB. Each sub-figure of Fig.~\ref{fig_const} contains four symbol clusters 
differentiated by different colors. Each cluster refers to one symbol transmitted to \textit{Rx1}. 
Within each cluster, there are four symbols, each corresponding to a symbol of \textit{Rx2}. 
For example, the blue colors denote symbol~$1$ of \textit{Rx1} distorted by four symbols of \textit{Rx2} and also polluted by noise. 
Thus, the DAE-ZIC generates distinguishable symbols at the receivers, and these symbols are nonuniform and adaptive to the interference intensity.
\begin{table}[htb]
	\caption{The actual equivalent channels used  in Fig.~\ref{fig_const}.}\label{tab_ch}
	\centering
	\begin{tabular}{c|ccc}\hline
		\vspace{-0.09in} & & &\\
		$\hat{\alpha}$ & $\bar{h}_{11}$ & $\bar{h}_{22}$ & $\bar{h}_{21}$\\ 
		\hline
		$0.5$   &  $1.14 - 0.06i$  &  $1.05 + 0.02i$     & $0.56 - 0.14i$   \\
		$1.0$ &  $1.01 + 0.11i$   &  $0.75 - 0.27i$    & $1.05 - 0.04i$   \\
		$1.5$  &  $1.24 - 0.13i$  &  $0.91 - 0.08i$ & $1.28 - 0.12i$  \\      \hline   
	\end{tabular}
\end{table}

It can be seen that the location and distribution of symbols  are different in each method.  
The constellations of \textit{Baseline-1} (left column) 
are very crowded and symbols are even overlapped when $\hat\alpha=1$ and $\hat\alpha=1.5$ because 4-QAM is directly applied.   \textit{Baseline-2} 
(middle column) rotates the constellation of \textit{Tx2}, which enlarges the space between 
symbols and thus helps reduce the decoding error. However, the imperfect CSI may cause high BERs, 
especially when $\hat\alpha=1$ and $\hat\alpha=1.5$. This is because the accuracy of CSI highly affects the optimization of the rotation angle.  
Differently, the DAE (right-column) can intelligently choose and adjust various scaled 
constellation types to avoid  constellation overlapping. 
When $\alpha=0.5$, the DAE-ZIC designs a parallelogram-shape constellation 
compared with the 
square-shaped constellations (4-QAM) in the baselines. When 
$\alpha=1$, both \textit{Tx1}  and \textit{Tx2} generate rectangular-shape constellations and 
inter-cross with each other. 
When $\alpha=1.5$, both \textit{Tx1} and \textit{Tx2} use PAM.
By adapting their constellations to the interference intensity, 
the two DAEs cooperate to avoid symbol overlapping. This is the main 
reason that the DAE-ZIC outperforms the baselines.

\begin{figure}[tbp]  
	\centering	
	\hspace{-0.08in}\subfigure[\textit{Baseline-1}, $\hat\alpha=0.5$]{
		\includegraphics[scale=.226, trim=0 0 0 0, 
		clip]{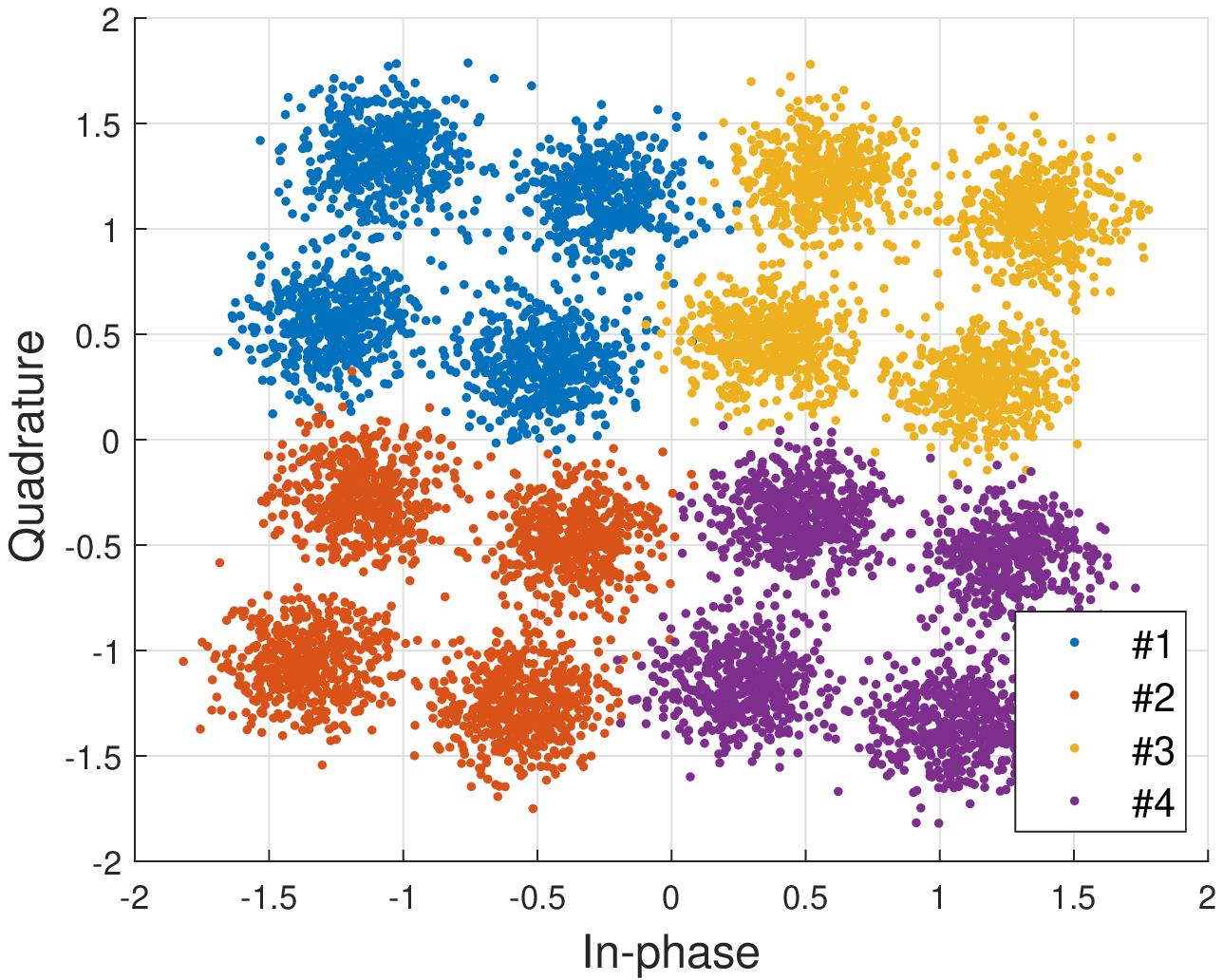}\label{fig_const_a1}}
	\hspace{-0.09in}\subfigure[\textit{Baseline-2}, $\hat\alpha=0.5$]{
		\includegraphics[scale=.226, trim=0 0 0 0, 
		clip]{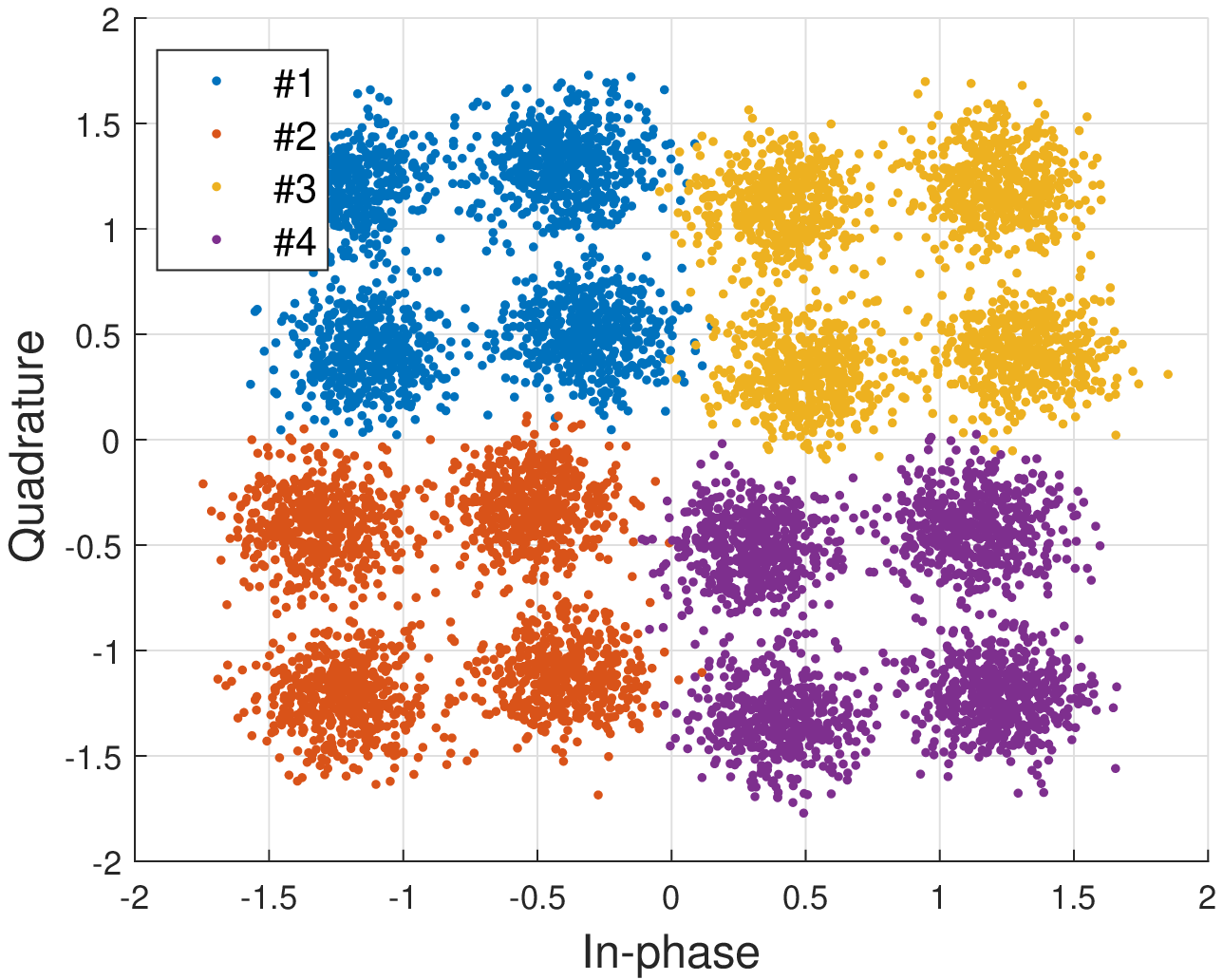}\label{fig_const_a2}}
	\hspace{-0.042in}\subfigure[DAE-ZIC, $\hat\alpha=0.5$]{
		\includegraphics[scale=.226, trim=0 0 0 0, 
		clip]{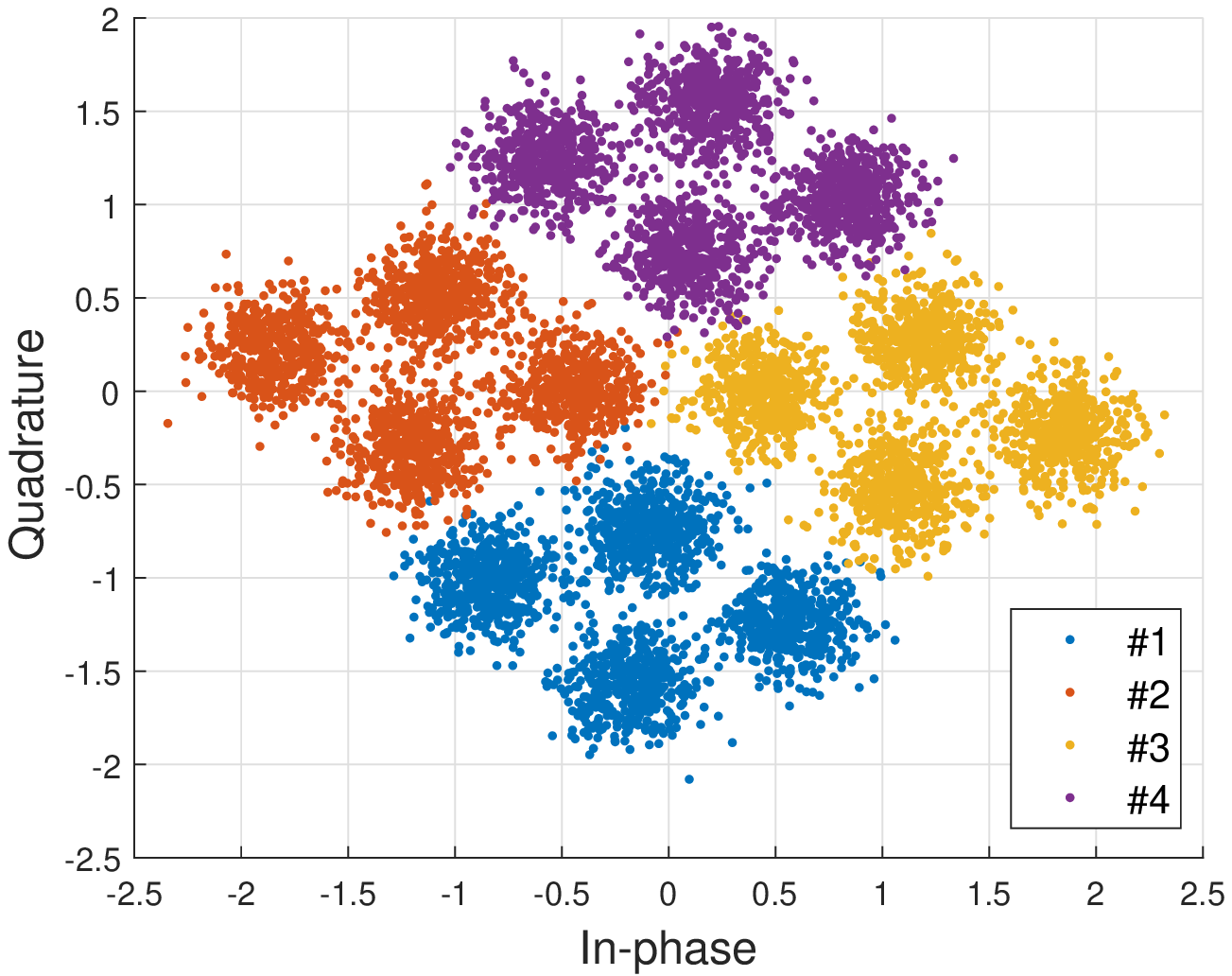}\label{fig_const_a3}}
	
	\hspace{-0.08in}\subfigure[\textit{Baseline-1}, $\hat\alpha=1$]{
		\includegraphics[scale=.224, trim=0 0 0 0, 
		clip]{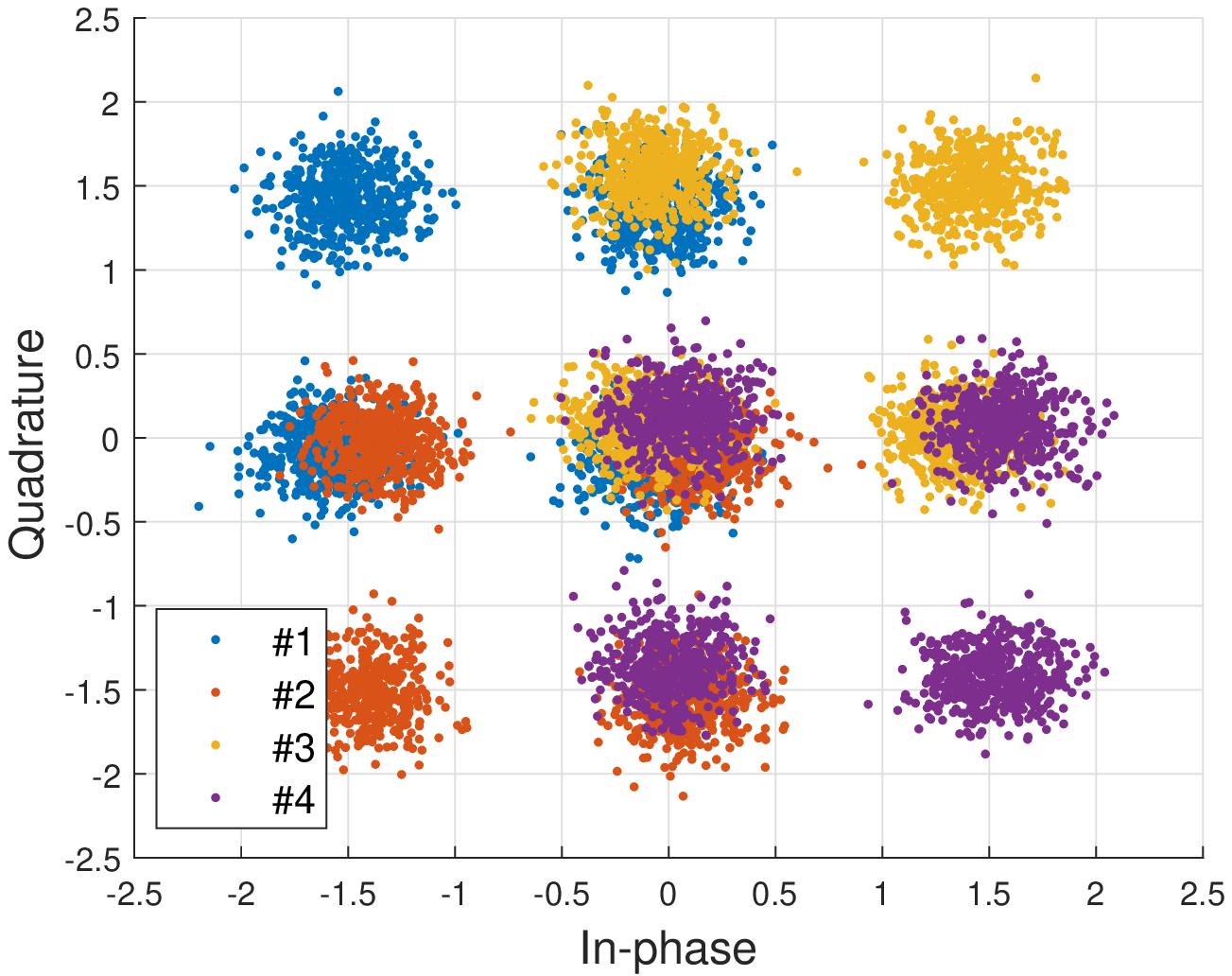}\label{fig_const_b1}}
	\hspace{-0.05in}\subfigure[\textit{Baseline-2}, $\hat\alpha=1$]{
		\includegraphics[scale=.224, trim=0 0 0 0, 
		clip]{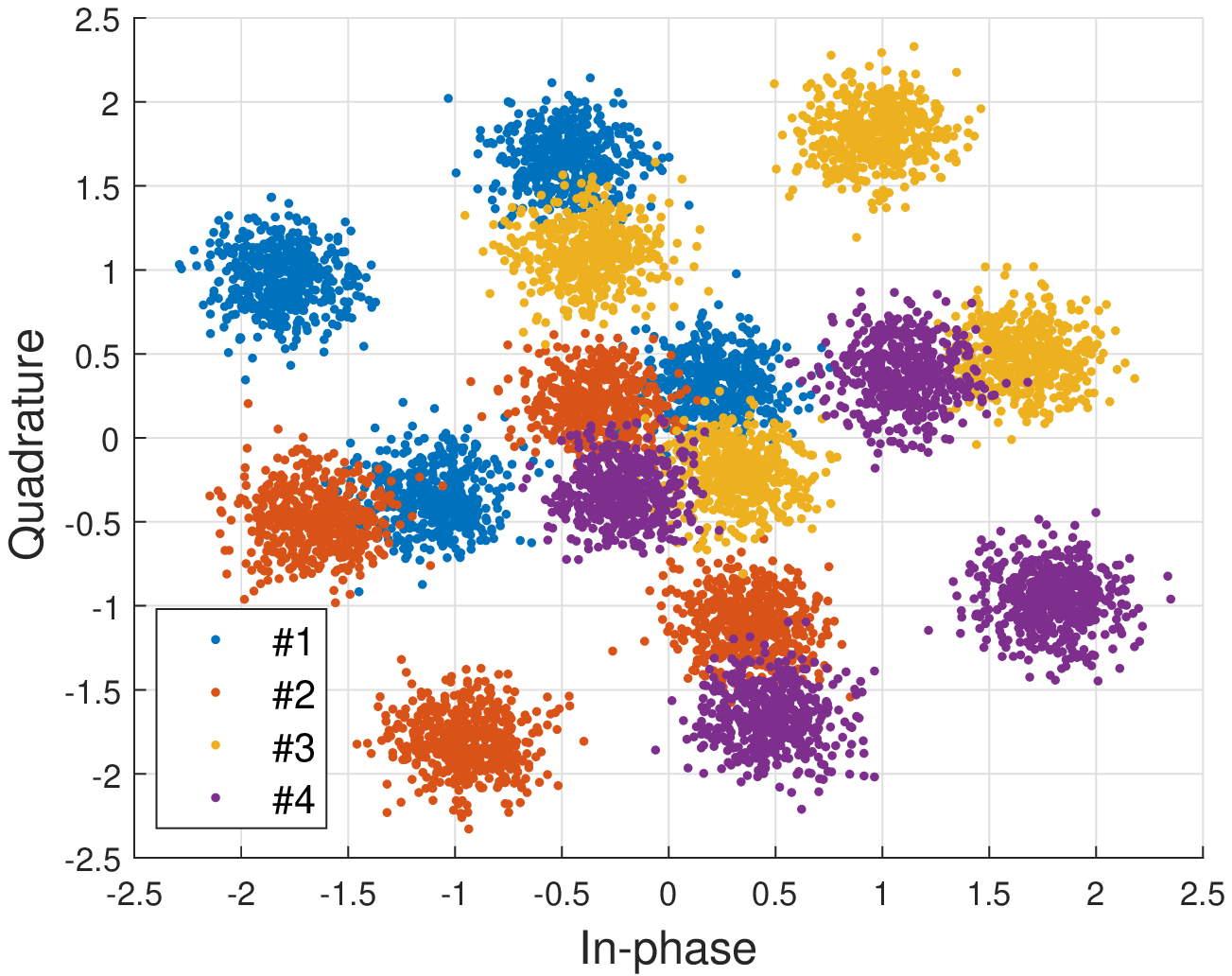}\label{fig_const_b2}}
	\hspace{-0.042in}\subfigure[DAE-ZIC, $\hat\alpha=1$]{
		\includegraphics[scale=.224, trim=0 0 0 0, 
		clip]{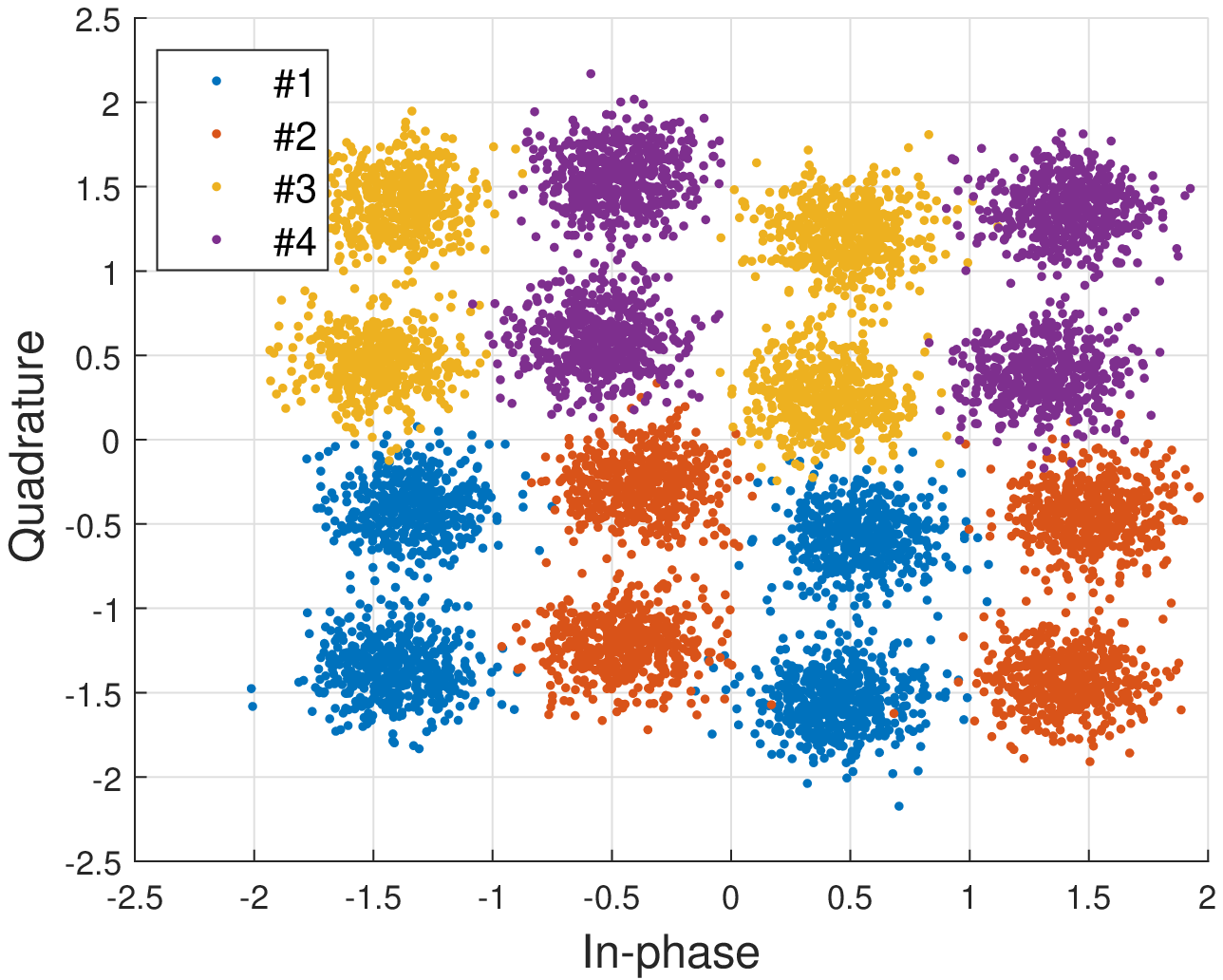}\label{fig_const_b3}}
	
	\hspace{-0.08in}\subfigure[\textit{Baseline-1}, $\hat\alpha=1.5$]{
		\includegraphics[scale=.224, trim=0 0 0 0, 
		clip]{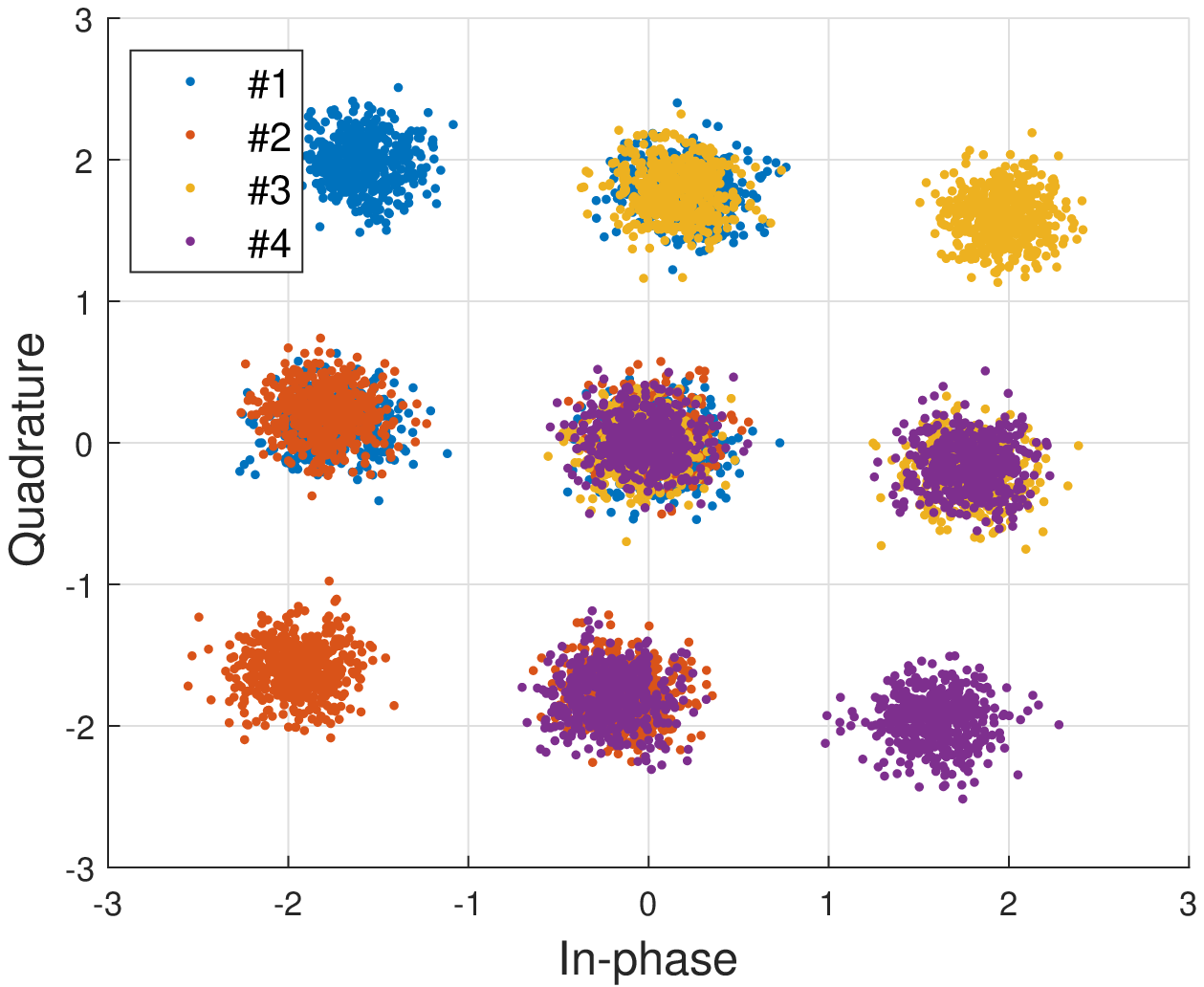}\label{fig_const_c1}}
	\hspace{-0.02in}\subfigure[\textit{Baseline-2}, $\hat\alpha=1.5$]{
		\includegraphics[scale=.224, trim=0 0 0 0,  
		clip]{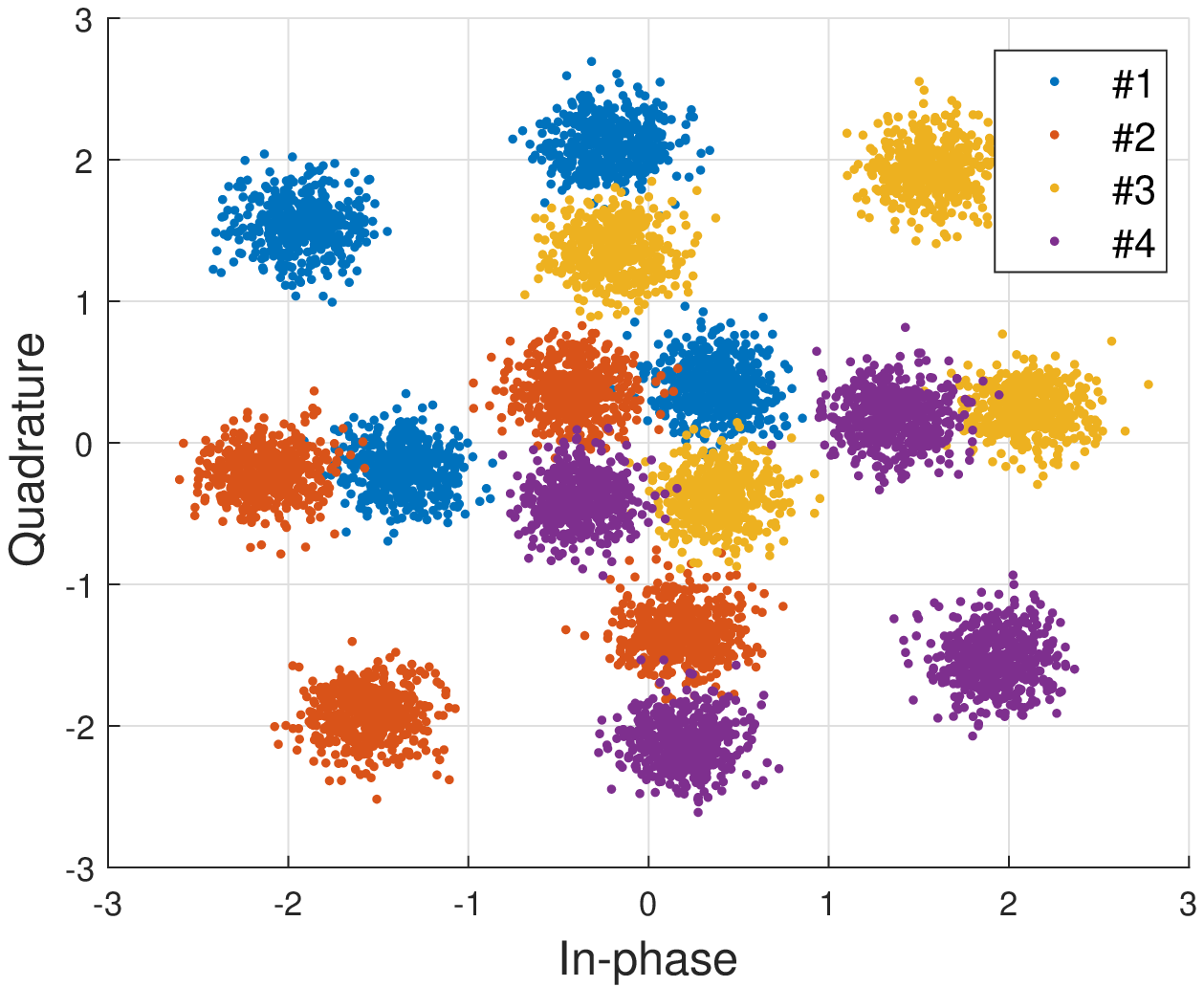}\label{fig_const_c2}}
	\hspace{-0.042in}\subfigure[DAE-ZIC, $\hat\alpha=1.5$]{
		\includegraphics[scale=.224, trim=0 0 0 0, 
		clip]{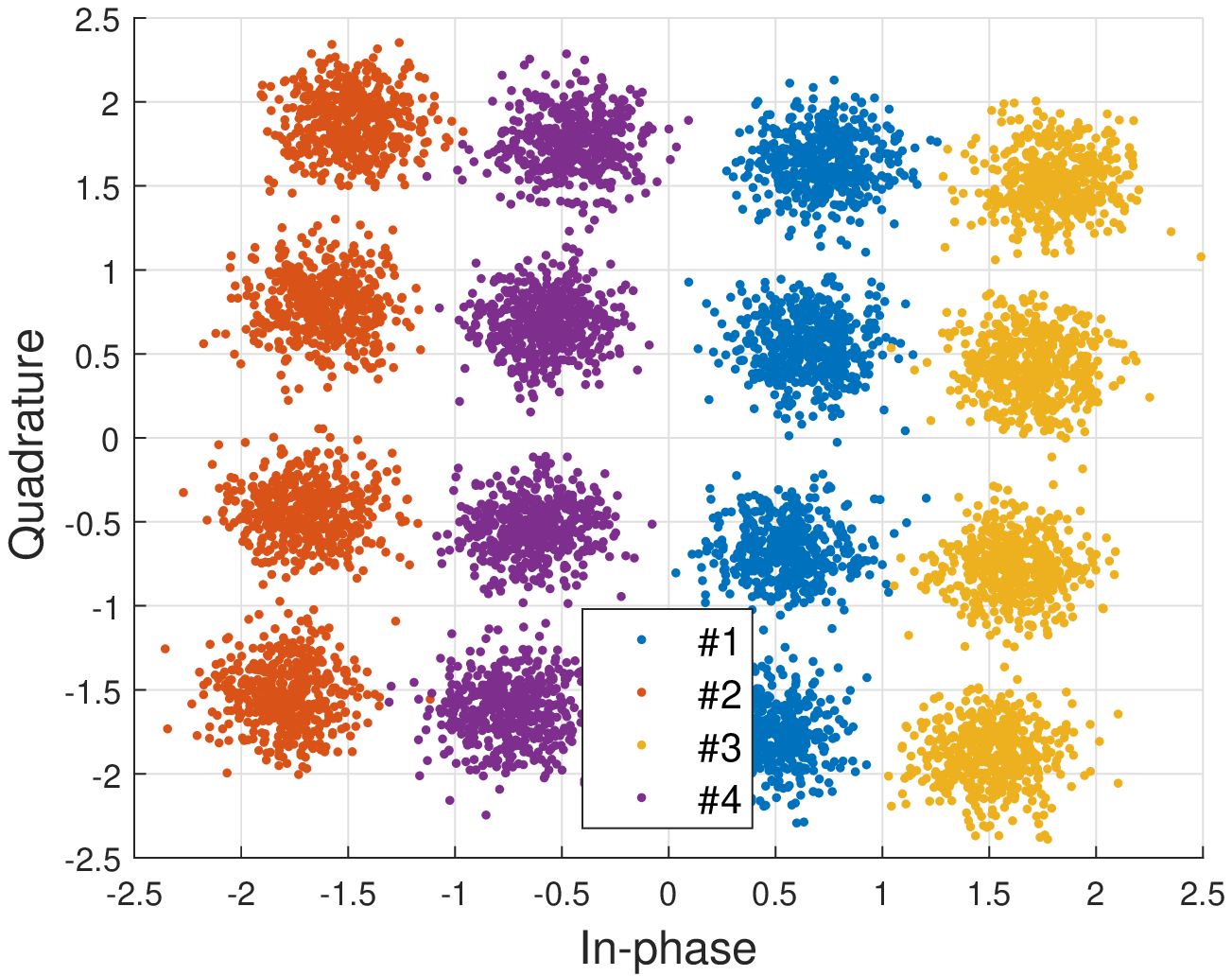}\label{fig_const_c3}}
	\caption{Constellations of the DAE-ZIC and the two baselines at \textit{Rx1}  with estimated 
	channels for three values of $\hat\alpha$.}
	\label{fig_const} 
\end{figure}
\begin{table}[tb]
	\caption{The percentage of BER reduction obtained by the  DAE-ZIC compared to \textit{Baseline-1} and \textit{Baseline-2} for different values of $\sigma_E^2$ and $N_s$. }\label{tab_impch_BER}
	\centering
	\begin{tabular}{l|l|cc|cc}
		\hline
		\multicolumn{2}{l|}{Compared to}  & \multicolumn{2}{c}{\textit{Baseline-1}} & 
		\multicolumn{2}{c}{\textit{Baseline-2}} \\ 
		\hline
		\multicolumn{2}{l|}{$N_s$}                    & 2          & 3          & 2          & 3          \\ \hline
		\multirow{3}{*}{$\sigma_E^2$}
		&$0$   &  75.77\%   &  44.29\%     & 44.43\% & 31.50\%   \\
		&$0.05$ &  55.40\%  &   38.97\%    & 39.12\% &  31.43\%   \\
		&$0.1$  &   48.83\%  &  35.81\%    & 41.41\% &    29.24\% \\ \hline        
	\end{tabular}
\end{table}

We evaluate the performance on two levels of estimation error, 
$\sigma_E^2\in\{0.05, 0.1\}$.  We test the 
BER over 15500 random channels. The interference intensity $\alpha$ is uniformly generated from $0$ to $3$ with step $0.1$, and SNR is $10$dB. The  BER reduction achieved by the 
 proposed DAE-ZIC compared to the baseline methods  is shown in Table~\ref{tab_impch_BER}.  
 The DAE-ZIC has a remarkable improvement in BER in all interference intensity and 
estimation error levels. The main improvement of the proposed 
DAE-ZIC comes from that it designs new, non-overlapping constellations based on 
the interference intensity.

\begin{figure*}[tbp] 
	\centering 
	\subfigure[${M}_1={M}_2=4$, $\alpha=0.5$]{
		\includegraphics[scale=.45, trim=0 0 0 0, 
		clip]{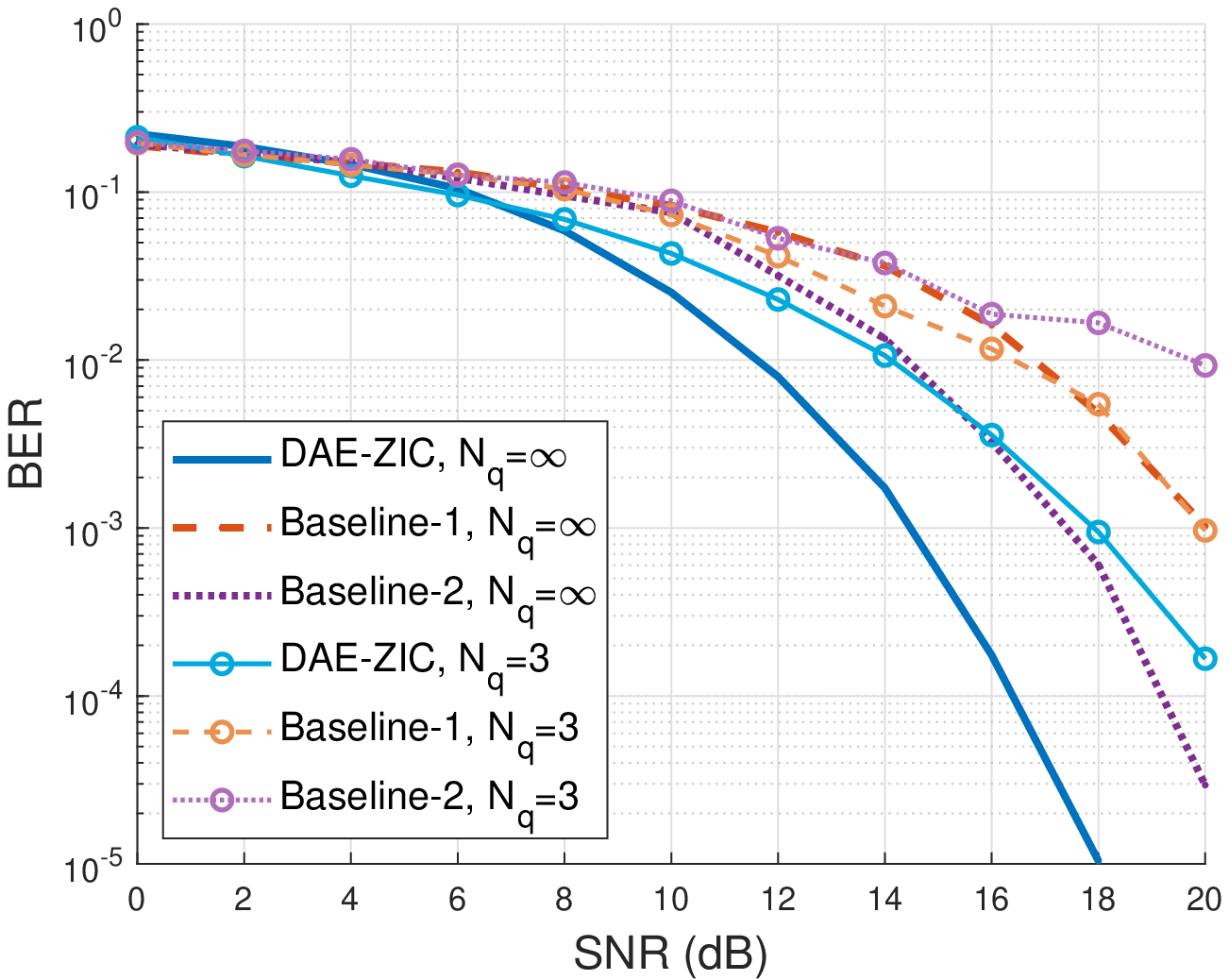}\label{fig_cmpSNRqt_a0.5}}\hspace{-2.1mm}
	\subfigure[${M}_1={M}_2=4$, $\alpha=1$]{
		\includegraphics[scale=.45, trim=0 0 0 0, 
		clip]{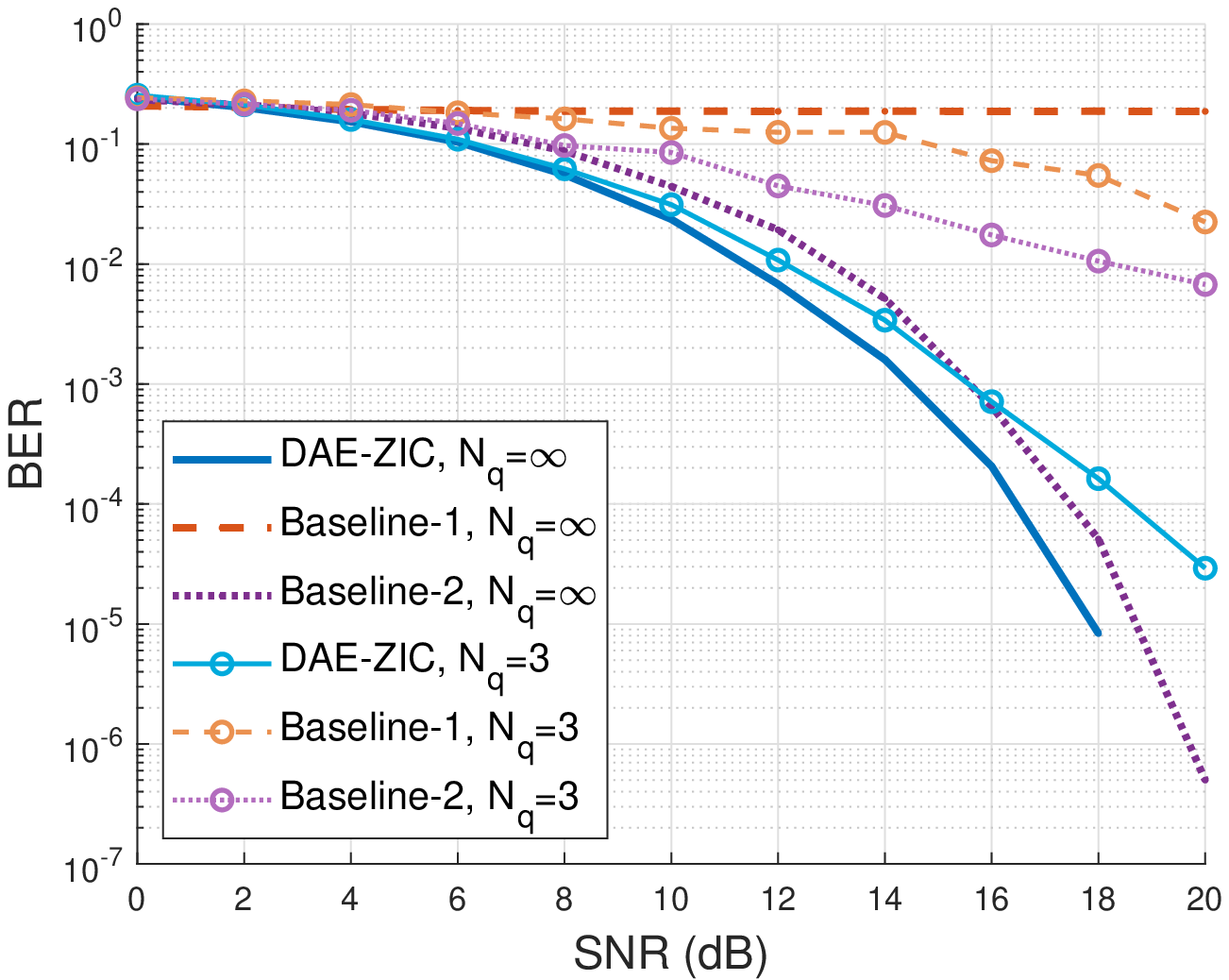}\label{fig_cmpSNRqt_a1}}\hspace{-2.1mm}
	\subfigure[${M}_1={M}_2=4$, $\alpha=1.5$]{
		\includegraphics[scale=.45, trim=0 0 0 0, 
		clip]{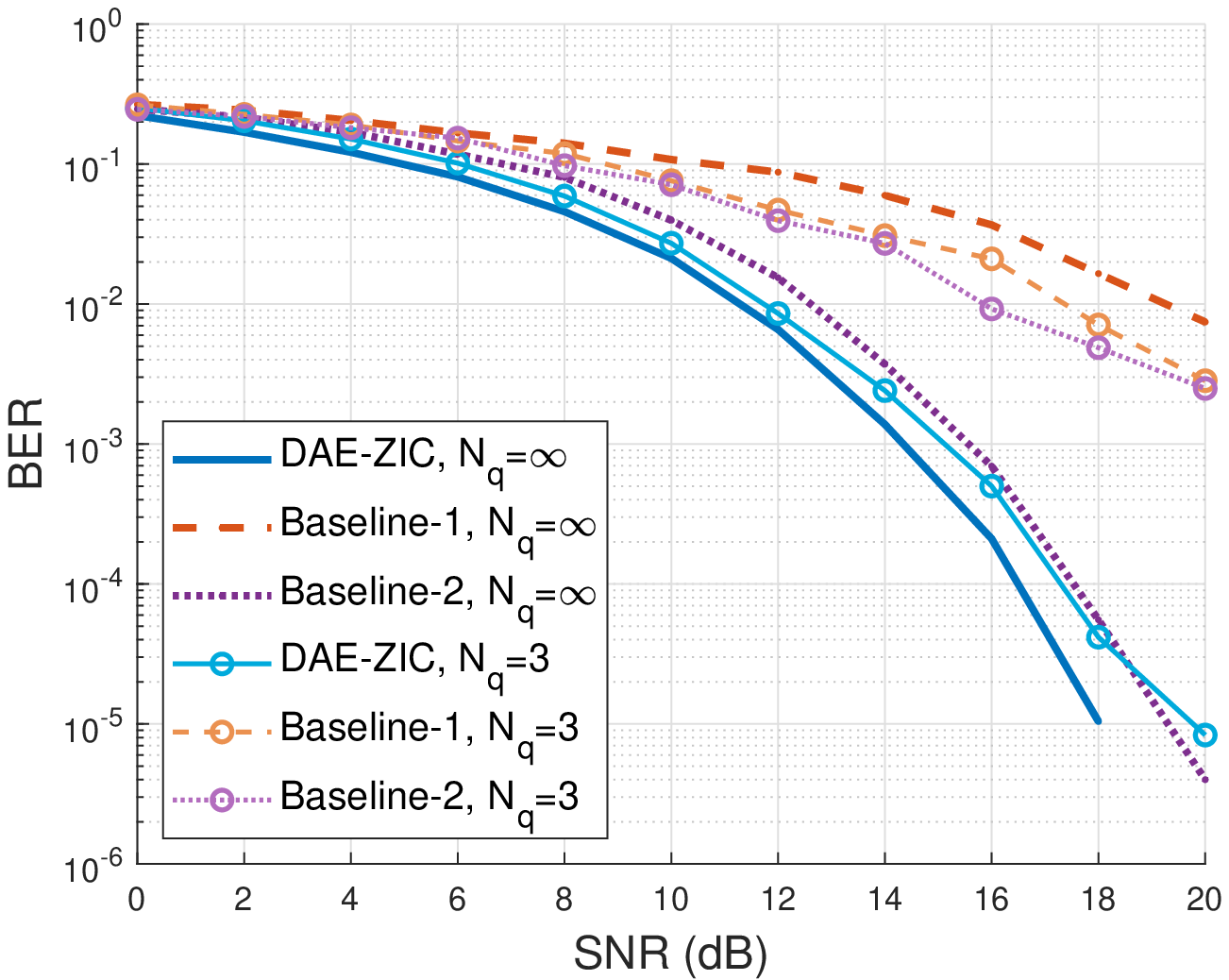}\label{fig_cmpSNRqt_a1.5}}\hspace{-2.1mm}
	\caption{The maximum (worst) BER between the two users (\textit{Rx1} and \textit{Rx2}) versus SNR with and without quantization error.}
	\label{fig_cmpSNRqt_1h0.1} 
\end{figure*}

\subsection{CSI with Feedback Quantization}
When channel estimation is perfect, the BER performance  with feedback quantization (with 
${N_q=3}$) and without quantization (${N_q=\infty}$)  is shown in 
Fig.~\ref{fig_cmpSNRqt_1h0.1}.  
The DAE-ZIC outperforms the baselines with and without feedback quantization.
It is seen that the quantization increases the BER of all methods. However, the performance 
degradation of the DAE-ZIC is much less than the other two methods. Especially, for $\alpha=1$ 
and $\alpha=1.5$), where the interference is strong, the DAE-ZIC outperform other methods by 
over two orders of magnitude when  $N_q=3$  and $\rm SNR=20$dB.
Interestingly, \textit{Baseline-1} performs  better for $N_q=3$ compared to $N_q=\infty$. The 
reason is that quantization of the angles  introduces an unintended rotation on \textit{Tx2} constellation, unintentionally acting like \textit{Baseline-2}. This then reduces the constellation overlapping and thus brings 
a better BER.

With high interference intensities, i.e,  $\alpha=1$ and $\alpha=1.5$, the BER of the DAE-ZIC 
with quantization ($N_q=3$) is only slightly degraded compared to the un-quantized case 
($N_q=\infty$).  
This is because, while \textit{Tx}s receive the quantized CSI, \textit{Rx1} knows CSI before and 
after quantization. Then, it  can to some extent correct the imperfectness in transmitters.  
Therefore, there is no dramatic degradation for the DAE-ZICs. On the other hand,  
\textit{Baseline-2} is more sensitive to the quantization error and thus a big gap of the BER 
happens between $N_q=\infty$ and $N_q=3$.} 
The reason is that  \textit{Baseline-2} only rotates the constellation in  
\textit{Tx2}, which highly depends on the phase shifted by the channels.

\section{Conclusion}\label{sec_con}
A DAE-based constellation design for the two-user ZIC in the presence of channel estimation error and quantization error has been proposed.  The DAE-ZIC  minimizes the  BER by 
jointly designing transmit and receive DAEs and optimizing them. A normalization layer is designed to meet the average power constraint.  The DAE-ZIC results in more efficient 
symbols to achieve a lower BER.  BER simulations verify the effectiveness of the proposed 
structure.  Our simulation results demonstrate the effectiveness of the proposed structure, and a comparison with two baseline models shows that the DAE-ZIC significantly outperforms both.

%
%

\begin{thebibliography}{10}
	
	\bibitem{carleial1975case}
	A.~Carleial, ``A case where interference does not reduce capacity (corresp.),''
	{\em IEEE Trans. Inf. Theory}, vol.~21, no.~5, pp.~569--570,
	1975.
	
	\bibitem{han1981new}
	T.~Han and K.~Kobayashi, ``A new achievable rate region for the interference
	channel,'' {\em IEEE Trans. Inf. Theory}, vol.~27, no.~1,
	pp.~49--60, 1981.
	
	\bibitem{etkin2008gaussian}
	R.~H. Etkin, D.~N. Tse, and H.~Wang, ``Gaussian interference channel capacity
	to within one bit,'' {\em IEEE Trans. Inf. Theory}, vol.~54,
	no.~12, pp.~5534--5562, 2008.
	
	\bibitem{motahari2009capacity}
	A.~S. Motahari and A.~K. Khandani, ``Capacity bounds for the {Gaussian}
	interference channel,'' {\em IEEE Trans. Inf. Theory},
	vol.~55, no.~2, pp.~620--643, 2009.
	
	\bibitem{wu2013linear}
	Y.~Wu, C.~Xiao, X.~Gao, J.~D. Matyjas, and Z.~Ding, ``Linear precoder design
	for {MIMO} interference channels with finite-alphabet signaling,'' {\em IEEE Trans. Commun.}, vol.~61, no.~9, pp.~3766--3780, 2013.
	
	\bibitem{foschini1974optimization}
	G.~Foschini, R.~Gitlin, and S.~Weinstein, ``{Optimization of two-dimensional
		signal constellations in the presence of Gaussian noise},'' {\em IEEE Trans. Commun.}, vol.~22, no.~1, pp.~28--38, 1974.
	
	\bibitem{goldsmith1997variable}
	A.~J. Goldsmith and S.-G. Chua, ``{Variable-rate variable-power MQAM for fading
		channels},'' {\em IEEE Trans. Commun.}, vol.~45, no.~10,
	pp.~1218--1230, 1997.
	
	\bibitem{barsoum2007constellation}
	M.~F. Barsoum, C.~Jones, and M.~Fitz, ``Constellation design via capacity
	maximization,''  in {\em Proc. IEEE Int. Symp. Inf. Theory}, pp.~1821--1825, 2007.
	
	
		\bibitem{vaezi2016simplified}
	M.~Vaezi and H.~V. Poor, ``Simplified {H}an-{K}obayashi region for one-sided
	and mixed {G}aussian interference channels,'' in {\em Proc.  IEEE Int. Conf. Commun. (ICC)}, pp.~1--6, 2016.
	
	
	\bibitem{knabe2010achievable}
	F.~Knabe and A.~Sezgin, ``Achievable rates in two-user interference channels
	with finite inputs and (very) strong interference,'' in {\em Proc. IEEE
		Asilomar Conf. Signals Syst. Comput. (ACSSC)},
	pp.~2050--2054, 2010.
	
	\bibitem{ganesan2012two}
	A.~Ganesan and B.~S. Rajan, ``Two-user {Gaussian} interference channel with
	finite constellation input and {FDMA},'' {\em IEEE Trans. Wirel. Commun.}, vol.~11, no.~7, pp.~2496--2507, 2012.
	
	\bibitem{o2017introduction}
	T.~O’Shea and J.~Hoydis, ``An introduction to deep learning for the physical
	layer,'' {\em IEEE Trans. Cogn. Commun. Netw.},
	vol.~3, no.~4, pp.~563--575, 2017.
	
	\bibitem{zhang2021svd}
	X.~Zhang, M.~Vaezi, and T.~J. O'Shea, ``{SVD}-embedded deep autoencoder for
	{MIMO} communications,'' in {\em Proc. IEEE Int. Conf. Commun. (ICC)}, pp.~1--6, 2022.
	
	\bibitem{song2020benchmarking}
	J.~Song, C.~H{\"a}ger, J.~Schr{\"o}der, T.~O'Shea, and H.~Wymeersch,
	``Benchmarking end-to-end learning of {MIMO} physical-layer communication,''
	in {\em Proc. IEEE Glob. Commun. Conf. (GLOBECOM)}, pp.~1--6,
	2020.
	

	\bibitem{erpek2018learning}
	T.~Erpek, T.~J. O'Shea, and T.~C. Clancy, ``Learning a physical layer scheme
	for the {MIMO} interference channel,'' in {\em Proc. IEEE Int. Conf. Commun. (ICC)}, pp.~1--5, 2018.
	
	\bibitem{wu2020deep}
	D.~Wu, M.~Nekovee, and Y.~Wang, ``Deep learning-based autoencoder for {M}-user
	wireless interference channel physical layer design,'' {\em IEEE Access},
	vol.~8, pp.~174679--174691, 2020.
	
	\bibitem{chen2005performance}
	Y.~Chen and C.~Tellambura, ``Performance analysis of maximum ratio transmission
	with imperfect channel estimation,'' {\em IEEE Commun. Lett.},
	vol.~9, no.~4, pp.~322--324, 2005.
	
	
	\bibitem{kramer2004review}
	G.~Kramer, ``Review of rate regions for interference channels,'' in {\em Proc.
		IEEE Int. Zurich Seminar Commun. (IZSC)}, pp.~162--165,
	2004.
	
		\bibitem{zhang2023wcncIC}
	X.~Zhang and M.~Vaezi, ``Deep autoencoder-based {Z}-interference channels,'' in
	{\em Proc. IEEE Wireless Commun. Netw. Conf. (WCNC)},
	2023.
	
\end{thebibliography}

\end{document}